# Sn-doped Bi$_{1.1}$Sb$_{0.9}$Te$_2$S, a bulk topological insulator with ideal properties

S. K. Kushwaha[1,*], I. Pletikosić[2,3], T. Liang[2], A. Gyenis[2], S. H. Lapidus[4], Yao Tian[5], He Zhao[6], K. S. Burch[6], Huiwen Ji[1], A. V. Fedorov[7], Ali Yazdani[2], N. P. Ong[2], T. Valla[3], and R. J. Cava[1]

[1]*Department of Chemistry, Princeton University, Princeton, NJ 08544, USA*

[2]*Department of Physics, Princeton University, Princeton, NJ 08544, USA*

[3]*Condensed Matter Physics and Materials Science Department, Brookhaven National Lab, Upton, New York 11973, USA*

[4]*X-ray Science Division, Advanced Photon Source, Argonne National Laboratory, Argonne IL 60439, USA*

[5]*Department of Physics, University of Toronto, Toronto, ON M5S 1A7, Canada*

[6]*Department of Physics, Boston College, Boston, MA 02467-3804, USA*

[7]*Advanced Light Source, Lawrence Berkeley National Laboratory, Berkeley, CA 94720, USA*

A long-standing issue in topological insulator research has been to find a material that provides an ideal platform for characterizing topological surface states without interference from bulk electronic states and can reliably be fabricated as bulk crystals. This material would be a bulk insulator, have a surface state Dirac point energy well isolated from the bulk valence and conduction bands, have high surface state electronic mobility, and be growable as large, high quality bulk single crystals. Here we show that this major materials obstacle in the field is overcome by crystals of lightly Sn-doped Bi$_{1.1}$Sb$_{0.9}$Te$_2$S (Sn-BSTS) grown by the Vertical Bridgeman method, which we characterize here via angle-resolved photoemission spectroscopy, scanning tunneling microscopy, transport studies of the bulk and surface states, and X-ray diffraction and Raman scattering. We present this new material as a bulk topological insulator that can be reliably grown and studied in many laboratories around the world.

Topological insulators (TIs) display a new state of quantum matter (1-5). In these 3-dimensional (3D) small band gap bulk "insulators", metallic two-dimensional (2D) spin-momentum locked Dirac fermions exist on the surfaces. These fermions, exhibiting a linear energy vs. wavevector dispersion relation, owe their existence and protection to the time-reversal-symmetry (6). 3D TIs have sparked wide interest in the research community due to both their fundamental scientific interest and the opportunities they offer for observing new electronic phenomena and properties that may be exploited for spintronics or other advanced electronic applications (7-13). Unfortunately, however, the transport properties of the actual materials involved are typically dominated by the bulk-electrons (5,14-16), making the true electronic properties of the surface states difficult to observe and exploit.

$Bi_2Se_3$ and $Bi_2Te_3$, with bulk band gaps of ~300 meV and ~150 meV respectively, are members of the larger $M_2X_3$ (M = Sb, Bi; X = S, Se, Te) tetradymite family and are considered to be fundamental materials platforms for the investigation of topological surface states (TSS). $Bi_2Se_3$ is a special case because it is a relatively simple material where the TSS Dirac point (DP) energy is well separated from the bulk valence band (BVB) and bulk conduction band (BCB) (5). However, even the best bulk $Bi_2Se_3$ crystals are heavily doped and metallic in nature (16), with their Fermi level in the BCB, and so the SS cannot be characterized independently of interference from the bulk states. In $Bi_2Te_3$ the surface state DP energy is below that of the top of the BVB and the crystals have large bulk carrier concentrations due to antisite defects, making this material unsuitable for probing the intrinsic character of the SS (17,18).

Thus, efforts have been underway since the inception of the field to find a material that is the ideal bulk TI for studying the intrinsic properties of the surface Dirac electrons (19-24). So far, the primary desirable characteristics for the ideal bulk material– to have a surface state DP well isolated in energy from the bulk states and a very low bulk carrier concentration, both of which would allow for the study of the Dirac electrons at very low energies without interference from the bulk states, high surface state mobilities, and reliable growability as high quality single crystals - have proven to be very difficult to satisfy in a single real bulk material. Until the current work, $Bi_2Te_2Se$ , with well-controlled stoichiometry and defect chemistry, has been considered as the best material from this perspective, but its DP energy is below the maximum energy of the BVB by ~60 meV, making it impossible to study the SS electrons near the DP without interference from bulk states (20,25-27); a major problem. An alternative material is the quaternary solid solution $(Bi_{1-x}Sb_x)_2(Se_{1-y}Te_y)_3$, (28-30), but the stoichiometry of that material is not fixed by any intrinsic thermodynamic materials characteristic such as chemistry or structure, making it unsuitable as a general materials platform for the study of TSS.

Here we describe the characterization of a new material that we argue satisfies the all the synthesis and electronic structure criteria to be an ideal bulk topological insulator. This material is based on modifications of the naturally occurring tetradymite mineral $Bi_2Te_{3-x}S_x$, and is derived by combining concepts in crystal chemistry, defect chemistry, strain-mediated chemical stability, the effect of atomic electronegativites on the absolute energies of bulk electronic states, and resonant level doping. The resultant is the stable, bulk insulating, high crystallinity vertical-Bridgeman-method grown single crystal material $Bi_{1.1}Sb_{0.9}Te_2S$, doped with a very small percentage of Sn. Large crystals with uniform excellent properties can be reproducibly grown. This material displays a surface state DP energy fully isolated from the BVB and BCB, very-low bulk carrier concentrations, strong domination of the conductance by surface currents below 150 K, and high quality quantum oscillations from the surface states. We have characterized the crystals by ARPES (Angle resolved photoemission) low temperature STM (scanning tunneling microscopy), electronic transport, x-ray diffraction and Raman spectroscopy and conclude that this new material possesses all the properties required to be an ideal TI. We argue that its growth will make the characterization of topological surface states widely available to researchers throughout the world.

Fig. 1 shows a schematic view of the design process that leads us to $Bi_{1.08}Sn_{0.02}Sb_{0.9}Te_2S$ (Sn-BSTS) as the ideal topological insulator. We start with $Bi_2Se_3$ and $Bi_2Te_3$, which are the fundamental tetradymite-type TI materials, but are non-ideal, as described above. To obtain bulk insulating TI crystals based on these tetradymites, two main approaches and their combination have been employed. In approach-**1**, starting from $Bi_2Se_3$, Te has been partially substituted for Se to form $Bi_2Te_2Se$ (BTS), where the electron-donating Se vacancies in $Bi_2Se_3$ are controlled by confining the Se to the middle layer in the quintuple layer (QL) structure, or, alternatively, by much the less well controlled random $Bi_2(Se_{1-x}Te_x)_3$ solid solution. In approach-**2**, the composition of Bi-site is adjusted by Sb-Bi partial substitution. This method has the advantage of also yielding materials whose DP is isolated from the bulk states, but the defect chemistry is again difficult to control. By using both approaches simultaneously, yielding tetradymites of the type $(Bi_{1-x}Sb_x)_2(Se_{1-y}Te_y)_3$, the energy of the SS DP can be placed in the bulk band gap and films and crystals can be grown with low bulk carrier concentrations at low temperature (28-30). The disadvantage of this material is the difficulty in controlling its composition and therefore its defect chemistry to grow high quality bulk crystals.

Here we develop an alternative material system based on incorporating S rather than Se in the middle layer of $Bi_2Te_2Se$. The incorporation of S in place of Se in the middle layer of the tetradymite QL sandwich decreases the absolute energy of the valence band due to the higher electronegativity of the S (31) and makes the surface DP isolated in energy from the bulk states. This is a parallel materials development path to approach **1**. However the formation of a thermodynamically stable compound of composition "$Bi_2Te_2S$" is not possible due to crystallographic strain from layer size mismatch, as was early described by Linus Pauling (32), and so a modification of the size of the Bi layer must be made through approach-**2** to obtain a stable material. Through experimentation we have found that that stable material composition is $Bi_{1.1}Sb_{0.9}Te_2S$. This is a result of chemical synergism: the incorporation of S in the middle layer requires a contraction of the metal layers (the ones containing Bi and Sb, Sb is smaller than Bi) for size matching and selects a Bi/Sb ratio near 1:1 for stability. This is a very strong crystal-chemical driving force for stability, and results in the optimal growth of this composition as large single crystal form from the melt. This material has a relatively low carrier concentration, but as a final step we employ the concept of resonant level doping, developed to optimize the properties of tetradymite thermoelectrics (24,33,34), to compensate for the last of the native defects present. We thus obtain a bulk-insulating ideal topological insulator through 1% Sn substitution for Bi, and develop the material $Bi_{1.08}Sn_{0.02}Sb_{0.9}Te_2S$ (Sn-BSTS). Our high quality crystals of Sn-BSTS, were grown by the VBT method (Fig. 1b). The crystalline perfection has been established by X-ray diffraction and Raman spectroscopy (35-37) (supplementary information).

The temperature-dependent bulk resistivity ($\rho$) plots (Fig. 1c) of three samples of Sn-BSTS from a 15 cm long single crystal boule, separated by a total of 8 cm along the boule, show that bulk Sn-BSTS is highly insulating. $\rho$ increases exponentially with decreasing T, attains a maximum value (~125 $\Omega$cm) at ~100 K, and starts decreasing with decreasing T. This decrease indicates that the resistivity of the insulating bulk material is short-circuited by the metallic TSS. Such a high $\rho$ is excellent for a tetradymite; at T<100 K the bulk resistivity must be substantially higher but it cannot be measured directly due to the dominance of the surface state conductance. The Log $\rho$ vs. $T^{-1}$ plots (Fig. 1d) exhibit an activated behavior with the activation energy, $\Delta \approx$ 165 meV. This $\Delta$ is very high and is consistent with half of the bulk band gap measured by ARPES (see below) ($\Delta_{transport}$ = $E_{g\ transport}$/2, so $E_{g\ transport}$ = 330 meV, $E_{gARPES}$ = 350 meV), an indication of the intrinsic nature of the semiconducting Sn-BSTS crystals. The Hall coefficient ($R_H$) (Fig. 1e) increases with decreasing T and attains very high values (~-5×10$^4$ cm$^3$/C) at low T. At high *T*, activated behavior can be seen. Some samples show a change in sign of $R_H$ from positive to negative on cooling, and for T <150 K, all samples have negative $R_H$, indicating *n*-type carrier conduction. The carrier densities (*n*) evaluated from $R_H$, at T < 150K, are as low as ~3×10$^{14}$ cm$^{-3}$ and almost T-independent. As the low-T transport is dominated by surface states, the bulk carrier concentrations may actually be substantially lower. The feature in $\rho$ near T~50 K, where surface state transport dominates the conductance, is reproduced in all of the crystals studied, and is discussed further below.

The low-energy band structure of Sn-BSTS was investigated by ARPES and is shown in Fig. 2; the observations are consistent with what is seen in other tetradymite-based materials. The pristine crystal surface shows conical states in the center of the surface Brillouin zone that are linearly dispersing at about 4 eVÅ from a DP that is about 120 meV below the Fermi level; $E_F$ is in the surface states, not the bulk states (Fig. 2a). The bands probed by ARPES were then electron doped through the deposition of K atoms on the surface, pushing the chemical potential into states that are not occupied in the pristine crystals. Consistent with expectations, the conical bands display no photon-energy dependence, a test of the $k_z$ dependence of their energy, and therefore represent surface states. The photon and *k* dependent ARPES spectra of the K-deposited crystals enabled the determination of the extrema of the bulk valence and conduction bands that form the energy gap in which the surface states reside, Fig. 2b. The DP was thus found to be isolated from the bulk bands, as the lowest energy bulk conduction band states appear 230 meV above it, and the highest bulk valence bands only reach energies of 120 meV below it. These findings are shown schematically in Fig. 2c. The bulk gap is thus found to be ~350 meV, consistent with the transport activation energy. $E_F$ falls in the surface states only, and the Fermi surface, shown in Fig. 2d, consists only of a small ring centered at the Γ point of the Brillouin zone, with the Fermi wave vector $k_F = 0.03$ Å$^{-1}$. Constant energy cuts through the band structure, Fig. 2d, show that the surface states, as one moves away from the DP, acquire some hexagonal warping, and are joined by the six petals of the bulk valence bands some 240 meV below the Fermi level in the pristine sample, or the circular bottom of the bulk conduction band when the chemical potential is raised by 110 meV. This is confirmed in our tunneling microscopy experiments.

The topographic image of the Te-terminated (001) surface of an Sn-BSTS crystal displays large atomically flat regions with atomic modulations corresponding to the rhombohedral crystal structure (Fig. 3a); the Fourier transform of this image reflects the hexagonal symmetry of the crystal surface (Fig. 3b). In addition to the Te atoms, several defects are observed (26). We probed the electronic structure on the surface of Sn-BSTS by using scanning tunneling spectroscopy (STS) (38-41). First, we measured the differential conductance, which is proportional to the local density of states, across the surface along a line of length of 90 nm (Fig. 3c). The spectra show a 'U' shape curve with a minimum at -120 meV and multiple inflection points. Since STS is sensitive to both TSS and bulk bands, to correctly identify the features in the spectra, the STS results are compared with the E-$k$ structure established by ARPES. We identify the observed changes in the STS spectra at around -200 meV and +120meV to be the indication of the edges of the bulk bands. The tunneling between these energies is dominated by the surface states. The minimum of the conductance corresponds approximately to the DP energy. The Dirac energy (120 meV below the Fermi energy) is in agreement with the ARPES data measured in different crystals from the same boule, showing the long range homogeneity of the material. The variation in the position of the minimum energy along the scan line is small (with 2.4 meV standard deviation), which suggests a homogeneous electronic structure at the nanometer length scale, in spite of the complex chemical system. This homogeneity at atomic length scales is consistent with the excellent longer-range crystalline homogeneity.

Scattering of the surface electronic states was examined by recording differential conductance maps at various energies (Fig. 3d, e). Although the conductance maps are strongly influenced by the Sb-Bi disorder, the Fourier transforms of the maps capture the signal corresponding to the scattering of TSS and reveal a linearly dispersing scattering cone between the BCB and BVB (Fig. 3f). To compare the STM and ARPES data, we consider a simple picture. First, we note that scattering between the opposite sides of the Dirac cone is prohibited by time-reversal symmetry. In the case of a single Dirac cone, the prohibition suppresses the amplitude of the largest scattering wave vectors, resulting in a scattering cone with a smaller diameter. In our STM results, we approximate the observed scattering cone as a manifold of vectors connecting regions of large density of states measured by ARPES. In this picture the boundary of the observed energy-momentum structure for a single linear dispersing band has half of the slope as measured by ARPES. The expected energy-momentum relation is indicated as a red line (Fig. 3f), in good agreement with our ARPES measurement.

To study the availability of the TSS for charge transport experiments and potential applications of Sn-BSTS in advanced devices, we performed transport measurements on crystals at 20 mK, under magnetic fields up to 18 T. Fig. 4a and b are the magnetoresistance (MR, $\rho_{xx}$ vs. H) and Hall ($\rho_{yx}$ vs. H) plots, respectively, measured by tilting H in the *x-z* plane (*z* is the direction of the surface normal and *x* is the direction of current flow in the basal plane). The $\rho_{yx}$ vs. H plots show that the charge transport is mainly due to electrons. Both $\rho_{xx}$ and $\rho_{yx}$ exhibit strong quantum oscillations at $\theta = 90$, which gradually die out as $\theta$ approaches 0. To visualize the oscillations and study their origin, the angular dependence of $\rho_{xx}$ and $\rho_{yx}$ are plotted as a function of $H^{-1}$ (Fig. 4c, d). The SdH oscillations are extracted from $\rho_{xx}$ and $\rho_{yx}$, and the resulting difference resistivity ($\Delta\rho$) plots are shown in Fig. 5a and b; the oscillations are as large as ~ 6%, with a low frequency. For a particular oscillation peak (n=5) the field value corresponding to the peak position (($\mu_0 H)^{-1}_{n=5}$) is plotted with tilt angle (Fig. 5c). It is clear that the peak shift is proportional to sin $\theta$, a reflection of the 2D TSS origin of the oscillations (16,18). The Fermi surface cross-section area ($S_F$) is calculated as 37 T. The 2D SS carrier density ($n_s$) is 8.93×10$^{11}$ cm$^{-2}$ and the SS carriers are *n*-type. Using the ARPES Fermi velocity ($v_F$ = 4eVÅ), the Fermi surface in this crystal is found to be 134 meV above the DP, consistent with ARPES and STM measurements, again an indication of the homogeneity of the material over macroscopic distances in the single crystal.

The TI crystals can be considered as having parallel conductance paths, with the measured electrical conductivity being a linear addition of bulk ($\sigma_{bulk}$) and surface conductances ($\sigma_s$), $\sigma = \sigma_{bulk} + \sigma_s$. With this in mind, the Sn-BSTS crystals allow for the temperature dependence of the surface state conductance to be characterized over a wide temperature range. Taking the data for the highest bulk resistivity crystal (Fig. 1b), $\sigma$ is plotted vs. $T^{-1}$ in Fig. 5d. At high $T$, $\sigma_{bulk}$ is dominant; it decreases rapidly as T decreases. Below 150 K, the conduction due to $\sigma_s$ dominates and results in an upturn in $\sigma$. Being conservative, and taking $\sigma$ for T < 100 K as clearly dominated by the surface transport, $\sigma_s$ can be extracted from $\sigma$; the resulting T dependence of $\sigma_s$ in the range 4 – 100 K is plotted in the inset of Fig. 5d. Similar to the raw $\rho$ and Hall results, $\sigma_s$ shows an unusual step at ~50 K. This feature in $\sigma_s$ may be purely due to SS behavior, or a structural phase transition in the underlying bulk crystal. Synchrotron diffraction analysis at low T (supplementary information) rules out the possibility of a bulk structural phase transition, and the STM images taken at low T do not show any sign of a surface reconstruction. Therefore this anomaly cannot reflect the presence of structural transitions either on the surface or in the bulk. We speculate that it may reflect the sensitivity of the conductance of the TSS on Sn-BSTS to the adsorption of a residual active gas (e.g. $H_2$ or $O_2$) in the experimental chamber, but more work will have to be done to determine its origin.

In summary, we have shown that highly insulating TI crystals of $Bi_{1.08}Sn_{0.02}Sb_{0.9}Te_2S$, with intrinsic semiconductor characteristics ($\Delta$~165 meV, and $n < 3\times10^{-14}$ $cm^{-3}$), can be successfully grown by VBT, with uniform properties over 8 cm or more of a single crystal boule. The bulk band gap is consistently estimated to be ~350 meV by spectroscopy and transport methods, and in the bulk gap only surface states are present. The DP is found to be completely exposed in the gap, ~120 meV above the valence band maximum and ~230 meV below the conduction band minimum. The chemical potential for these crystals falls in the surface states only, at an energy of ~120 meV above the DP. The quantum oscillations from the surface states are high quality. The SS carriers in Sn-BSTS are *n*-type, which are preferable for charge transport experiments and potential spintronics applications. Our results lead us to conclude that Sn-BSTS is an ideal tetradymite-type bulk TI, and that it will be highly advantageous for use in many laboratories world-wide for future experimental studies of topological surface states, especially for the characterization of surface state quantum transport very close to the Dirac point.

**Methods**

The bulk single crystals were grown by the vertical Bridgman technique (VBT). The required amounts (according to the formula: $Bi_{1.08}Sn_{0.02}Sb_{0.9}Te_2S$) of elements with high purities (5N) were placed in C-coated bottom-pointed quartz ampoules of inner diameter 4 mm. The as-received high purity starting materials (Bi, Sb and Te) were further purified by heating under vacuum of ~ $10^{-5}$ mbar, at 950 °C in the presence of C in sealed quartz tubes. The crystal growth ampoules were sealed under a vacuum of ~ $10^{-5}$ mbar. The crystal growth was performed at ~700 °C with a translation rate of ampoule of 1 mm/hr through the crystallization zone (24), and ~15 cm long crystal boules were obtained. The crystals were studied for phase identification on a laboratory X-ray diffractometer (Bruker eco D8 Advance) equipped with Cu K$\alpha$ source and LYNXEYE XE detector. To check the long-range homogeneity of the crystal structure we also performed Raman microscopy at ambient temperature using 10 μW from a 532nm excitation laser, focused to 1μm spot. Synchrotron powder X-ray diffraction was performed on beamline 11-BM at the advanced photon source to probe the bulk crystal structure at low temperature. The bulk resistivity and carrier density were measured in four-probe and Hall configurations on a Quantum Design Physical Properties Measurement System (PPMS). The electrical contacts were made on the (001) surface through Pt-wires with silver paint. The electronic band structure was established by ARPES using a Scienta SES-3000 electron spectrometer set to the resolution of 12 meV and 0.1° at beamline 12.0.1 of the Advanced Light Source (ALS). Photon energies were in the range of 30 to 100 eV. Samples were cleaved at 15 K in ultrahigh vacuum of $5\times10^{-9}$ Pa and all the data were collected at 15 K. The STM measurements were performed on the (001) crystal surface in the temperature range of 30 K to 90 K, on a home-built variable temperature STM. Angular dependent magnetoresistance (MR) and Hall data were obtained in a dilution refrigerator at T = 20 mK at applied fields up to $\mu_0H$ = 18 T, using standard six contact measurements. Shubnikov-de Haas (SdH) quantum oscillations were obtained by subtracting the background from the data.

**Acknowledgements**

This research was supported by the ARO MURI on topological insulators, grant W911NF-12-1-0461, ARO grant W911NF-11-1-0379, and the MRSEC program at the Princeton Center for Complex Materials, grant NSF-DMR-1005438. The ARPES experiments were performed under the LBNL and BNL grants DE-AC02-05CH11231 and DE-SC0012704. The ALS is operated by the US DOE under Contract No. DE- AC03-76SF00098. . This research used resources of the Advanced Photon Source, a U.S. Department of Energy (DOE) Office of Science User Facility operated for the DOE Office of Science by Argonne National Laboratory under Contract No. DE-AC02-06CH11357. The Raman experiments were conducted with support from the National Science Foundation (grant DMR-1410846).

**Author Contributions**

S.K.K., H.J. and R.J.C designed the research. S.K.K. synthesized compound, performed optimized crystal growth, X-ray and bulk electronic measurements. I.P. and T.V. performed ARPES studies, T.L. and N.P.O. carried out and interpreted the transport measurements, A.G. and A.Y. performed the low temperature STM measurements, Y.T., H.Z. and K.S.B. performed the Raman experiments, and S.L. performed the low temperature X-ray diffraction measurements. S.K.K. & R.J.C. wrote the manuscript with input from all the coauthors.

**Competing financial interests:** Authors do not have any competing financial interest.

* kushwaha@princeton.edu

**Figure Legends:**

**Figure 1. Evolution of tetradymite-type TIs and bulk electronic properties ($\rho$ and $n$) of Sn-BSTS. a**, Schematic representation of the evolution of $Bi_2Se_3$, ($M_2X_3$) type TIs, through the two approaches, **1** & **2**. The solid and zebra lines show the single and multi-element layers, respectively. The left inset is a representation of a quintuple layer (QL) of general compounds and the right shows a QL of the titled crystal. **b**, the crystal boule obtained by vertical Bridgman technique (VBT), an orange line bar represents 1 cm scale. **c**, The $\rho$ vs. T plots for three freshly cleaved representative samples, A, E and G, respectively taken from the segments at 1, 6 and 10 cm from crystal boule. **d**, The log $\rho$ vs. 1/T plots. The straight lines are exponential fits to the high T data points. **e**, The Hall coefficient ($R_H$) vs. T plots for the same crystals. The dotted lines are guides to the eye showing the change of sign of $R_H$. The Hall data were recorded by applying magnetic field of 1 T normal to the (00$l$) sample surface, and the current flow was in the basal plane.

**Figure 2. Probing the electronic structure of Sn-BSTS through ARPES measurements on the (001) crystal surface. a**, The ARPES-determined band dispersion for $k_y$ along M–Γ–M measured at 30eV photon energy, at 15 K. The bands were electron doped through deposition of potassium (K) to allow for the determination of the energy of the conduction band minimum. **b**, Excerpts from the excitation energy dependence study of the photoemission from the electron doped bands reflect the perpendicular momentum $k_z$ positions of the bulk valence band (BVB) and conduction band (BCB) extrema. The surface states are unaffected by the change of $k_z$. **c**, A schematic for the electronic structure of Sn-BSTS in the vicinity of the Fermi energy derived from the ARPES data. Dashed lines mark the bulk band gap region, two crossed red lines correspond to the surface states, and the turquoise line shows the position of the Dirac point, DP. **d**, Constant-energy ARPES maps, referenced relative to the Fermi level $E_F$ for the pristine sample (on the right) or the Dirac point $E_{DP}$ for the K-deposition-generated electron doped sample (on top), showing a few critical points in the low energy band structure of Sn-BSTS.

**Figure 3. Mapping of lattice topograph, electronic conductance, quasi-particle interference and 2D surface states in Sn-BSTS by STM. a**, Real space topographic image of the (00l) surface of an Sn-BSTS crystal at $V_{bias}$ = - 600 mV set-point bias voltage and $I$ = 60 pA tunneling

current. **b**, Fourier transform of the topographic image revealing the Bragg peaks corresponding to the hexagonal surface structure. **c**, The differential conductance (*dI/dV*) along a line shows only small fluctuations in the electronic surface structure ($V_{bias}$ = - 600 mV and $I$ = 60 pA). The red line indicates the minima of the spectra, while the brown line displays the average spectrum along the line. Arrows point to the approximate positions of the edges of the bulk valence and conduction bands (BVB and BCB) and the position of the Dirac point (DP). **d**, **e**, Real space differential conductance (dI/dV) maps recorded at 60 and -300 meV with respect to the Fermi level. **f**, Fourier-transform scanning tunneling spectroscopy (STS) maps at different energies, reveal the structure of the quasiparticle scattering on the surface of the crystal. The images were obtained from symmetrized (considering the mirror and rotational symmetry of the underlying crystal). Fourier transforms of *dI/dV* conductance maps of size of 1320 Å x 1320 Å ($V_{bias}$ = - 500 mV and $I$ = 80 pA). The red lines with a slope of 2 eV/Å$^{-1}$ correspond to the expected dispersion relation that should be seen by STM, based on the ARPES measurements (figure 3, $v_{ARPES}$ = 4 eV/Å$^{-1}$). All measurements at 30 K.

**Figure 4. Magnetoresistance, Hall, and quantum oscillation characterization of the charge transport in Sn-BSTS. a**, Magnetoresistance and **b**, Hall plots recorded at 20 mK and their dependence on the angle (θ) between H and the current in the sample. **c**, $\rho_{xx}(H, \theta)$ and **d**, $\rho_{yx}(H, \theta)$ vs. $(\mu_0 H)^{-1}$ plots as a function of θ.

**Figure 5. The Shubnikov-de Hass (SdH) oscillations from the surface states in Sn-BSTS at 20mK**. **a** and **b**, The SdH oscillations respectively extracted from $\rho_{xx}$ and $\rho_{yx}$ for θ=90°. **c**, The θ dependence of $(\mu_0 H)^{-1}{}_{n=5}$ peak of SdH oscillations for $\rho_{xx}$. **d**, The total conductivity, $\sigma$ (= $\sigma_{bulk}$ + $\sigma_s$, where $\sigma_{bulk}$ and $\sigma_s$ are bulk and surface conductivities, respectively) plotted as a function of inverse temperature. The inset shows the $\sigma_s$ vs. T plot for T < 100K, extracted from $\sigma$.

Figure 1.

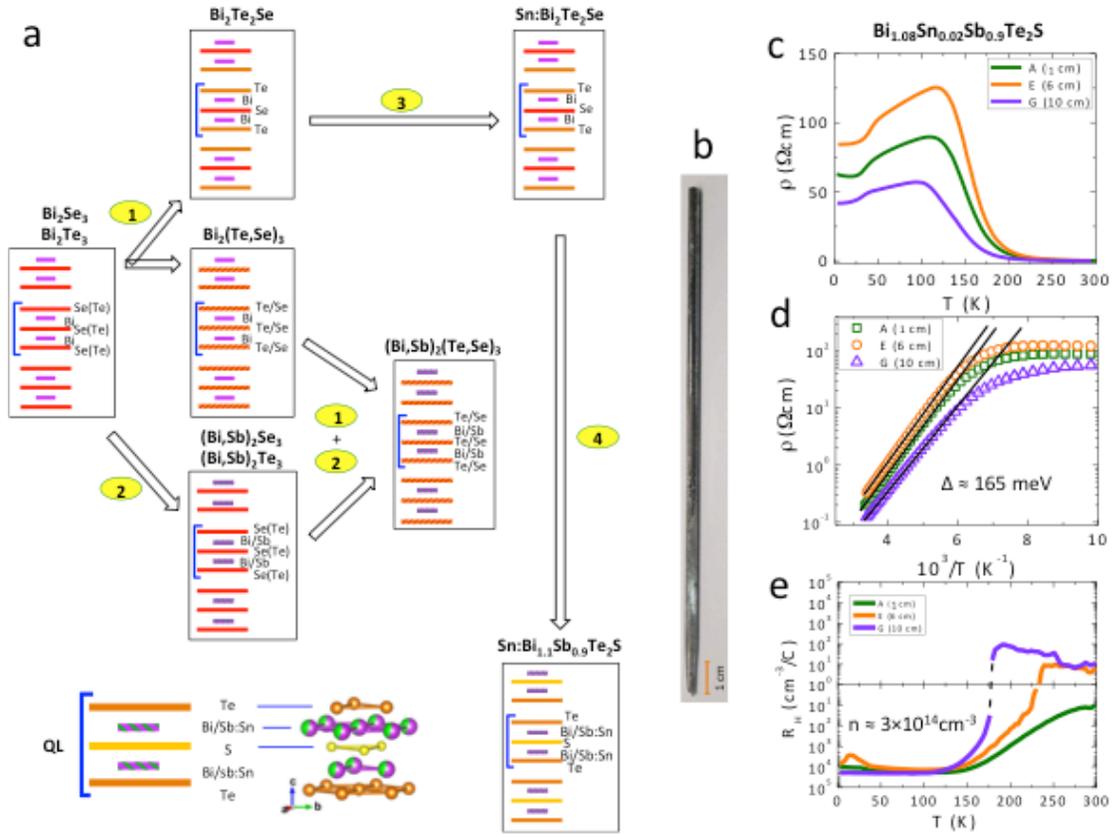

Figure 2.

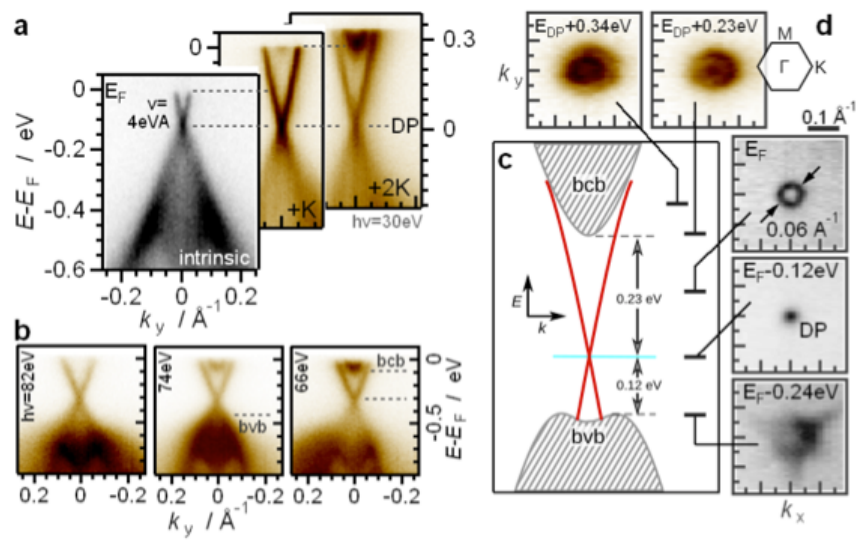

Figure 3.

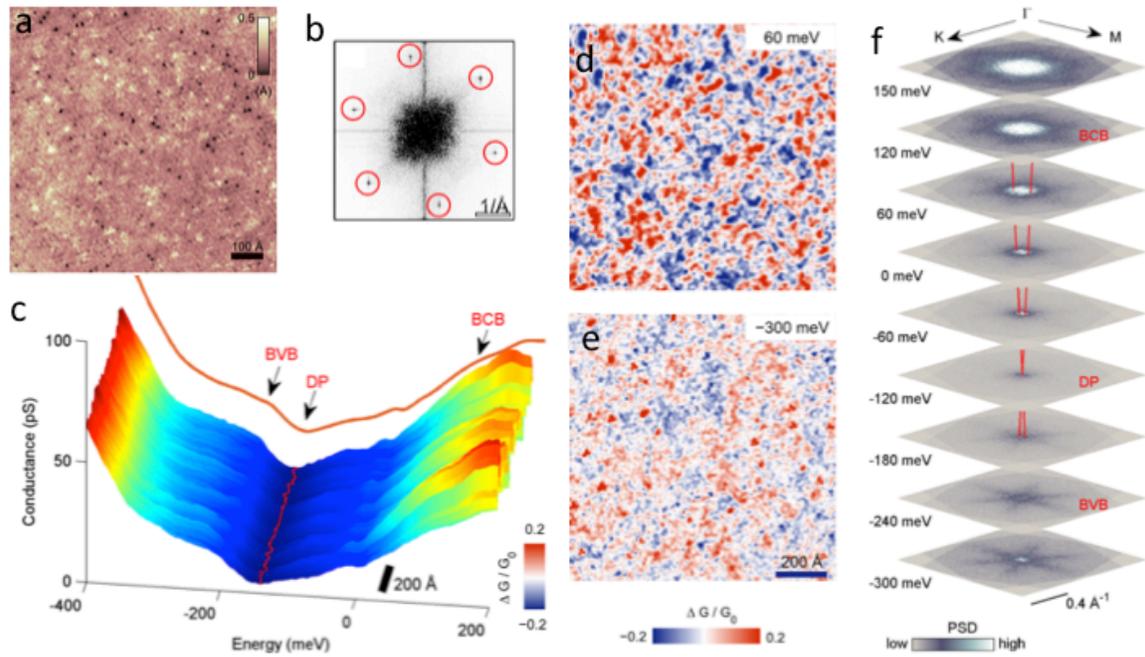

Figure 4.

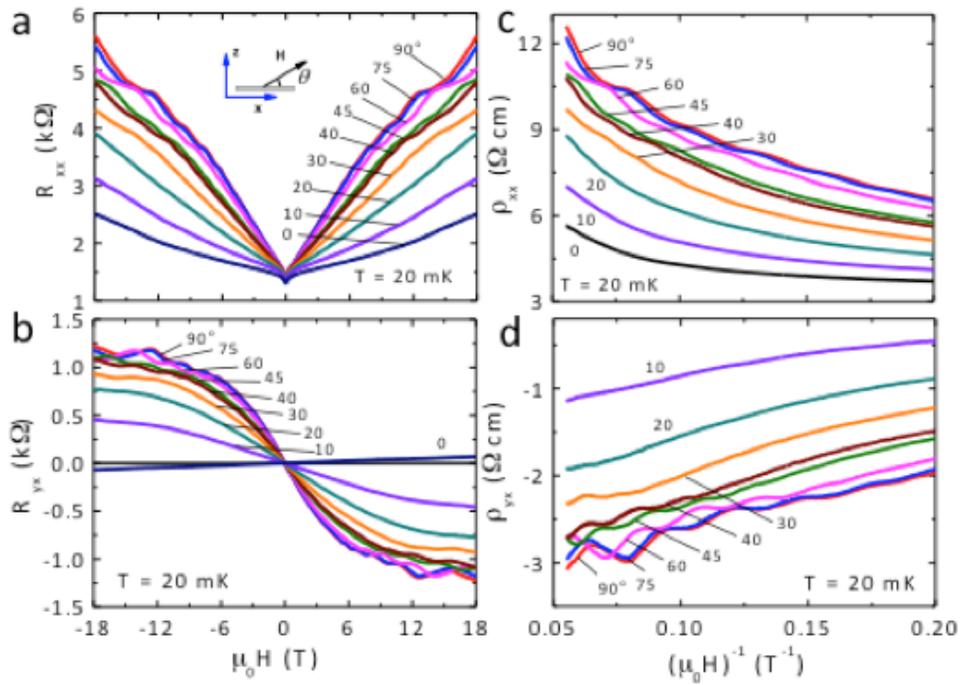

Figure 5.

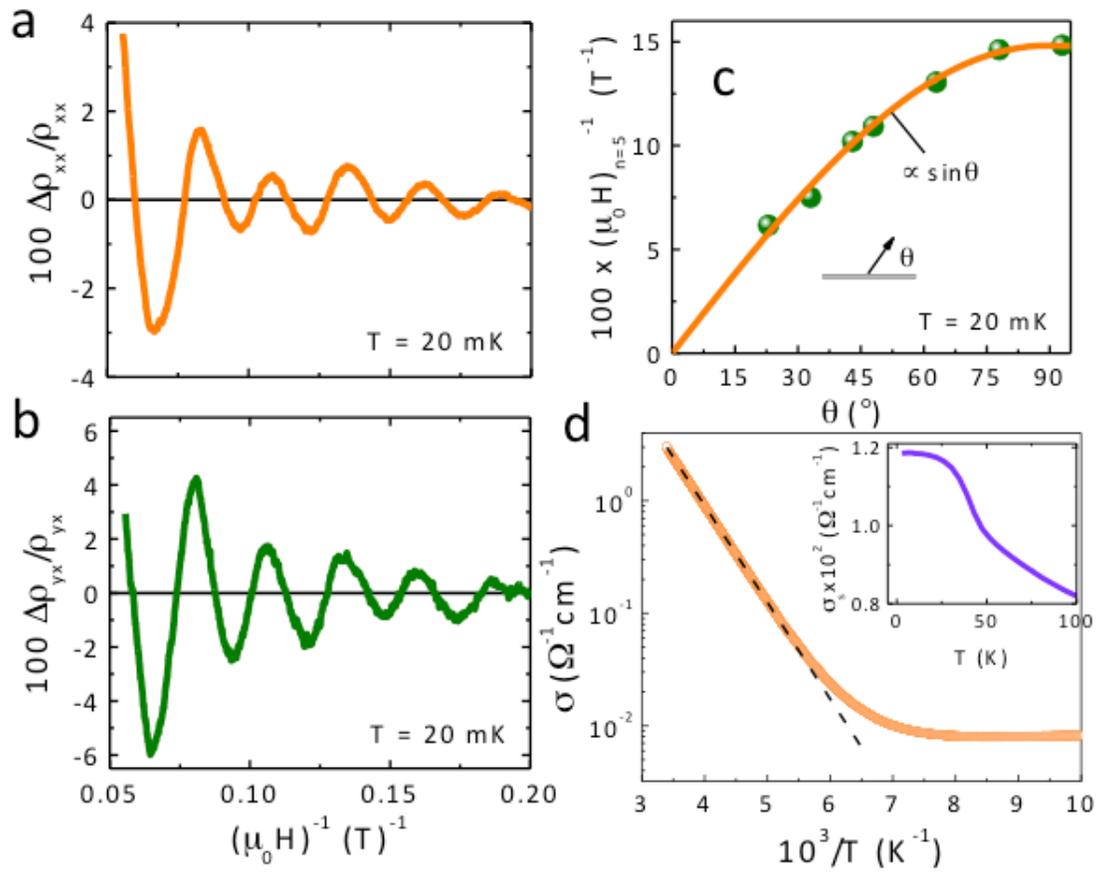